\documentclass[aps,prb,final,twocolumn,superscriptaddress]{revtex4-1}

\pdfoutput=1

\usepackage{amsmath}

\usepackage{amsfonts}
\usepackage{amsmath}
\usepackage{amssymb}
\usepackage{mathtools}
\usepackage{physics}

\usepackage{framed}
\usepackage{braket}
\usepackage{float}

\usepackage{cmmib57}

\usepackage{graphicx}
\usepackage{epsfig}
\usepackage{subcaption}

\usepackage{color}

\newcommand{\iu}{\mathrm{i}}
\newcommand{\e}{\mathrm{e}}
\newcommand{\vrv}{\boldsymbol{r}}
\newcommand{\vrvt}{\boldsymbol{r}_p}
\newcommand{\rt}{r_p}
\newcommand{\vh}{\boldsymbol{h}}
\newcommand{\vB}{\boldsymbol{B}}
\newcommand{\vk}{\boldsymbol{k}}
\newcommand{\vkt}{\boldsymbol{k}_p}
\newcommand{\kt}{k_p}
\newcommand{\vn}{{\boldsymbol{n}}}
\newcommand{\vx}{{\boldsymbol{x}}}
\newcommand{\vy}{{\boldsymbol{y}}}
\newcommand{\vz}{{\boldsymbol{z}}}

\newcommand{\ve}{{\boldsymbol{e}}}

\newcommand{\tpsi}{\tilde{\psi}}

\newcommand{\txi}{\tilde{\xi}}
\newcommand{\txio}{\tilde{\xi}_1}

\newcommand{\Oo}{\tilde{\Omega}_0}
\newcommand{\Os}{\tilde{\Omega}_u}

\newcommand{\Dpm}{\tilde{D}^{(\pm)}}
\newcommand{\Dm}{\tilde{D}^{(-)}}
\newcommand{\Dp}{\tilde{D}^{(+)}}
\newcommand{\Ds}{\tilde{D}^{(u)}}

\newcommand{\epm}{\epsilon}
\newcommand{\pto}{W}

\newcommand{\vgx}{v_x}

\newcommand{\up}{\xi_p}
\newcommand{\um}{\xi_m}
\newcommand{\Pm}{P_m}
\newcommand{\Fx}{F}

\newcommand{\dg}{^\mathrm{o}}

\newcommand{\be}{\begin{equation}}
\newcommand{\ee}{\end{equation}}
\newcommand{\bea}{\begin{eqnarray}}
\newcommand{\eea}{\end{eqnarray}}

\begin{document}



\title{Scattering of spin waves by a Bloch domain wall: effect of the dipolar interaction}



\author{Victor Laliena}
\email[]{laliena@unizar.es}
\affiliation{
  Departamento de Matem\'atica Aplicada, Universidad de Zaragoza,
  C/ Mar\'ia de Luna 3, E-50018 Zaragoza, Spain
}

\author{Athanasios Athanasopoulos}

\author{Javier Campo}
\email[]{javier.campo@csic.es}
\affiliation{Instituto de Nanociencia y Materiales de Arag\'on (CSIC - Universidad de Zaragoza)
and Departamento \\de F\'{\i}sica de Materia Condensada, Universidad de Zaragoza 
C/ Pedro Cerbuna 12, E-50009 Zaragoza, Spain}

\begin{abstract}
  It is known that a Bloch domain wall in an anisotropic ferromagnet is transparent to spin waves.
  This result is derived by approximating the dipolar interaction between magnetic moments by an effective
  anisotropy interaction. In this paper we study the the scattering of spin waves by a domain wall taking
  into account the full complexity of the dipolar interaction, treating it perturbatively in the distorted wave Born
  approximation.
  Due to the peculiarities of the dipolar interaction, the implementation of this approximation is not
  straightforward. The difficulties are circumvented here by realizing that the contribution of the dipolar interaction
  to the spin wave operator can be split into two terms: i) an operator that commutes with the spin wave operator in absence
  of dipolar interaction, and ii) a local operator suitable to be treated as a perturbation in the distorted wave Born approximaton.
  We analyze the scattering parameters obtained within this approach. It turns out that the reflection coefficient does
  not vanish in general, and that the transmitted waves suffer a lateral shift even at normal incidence.
  This lateral shift can be greatlty enhanced by making the spin wave go through an array of well separated domain walls.
  The outgoing spin wave will no be attenuated by the scattering at the domain walls since the reflection coefficient
  vanishes at normal incidence. This effect may be very useful to control the spin waves in magnonic devices.
\end{abstract}

\pacs{111222-k}

\maketitle


\section{Introduction}
\label{sec:intro}

Replacing electric currents by spin waves as a means to transfer and manipulating information in information technology
devices is currently seen as an alternative that might be revolutionary, due to the ultralow power consumption involved
in the propagation of spin waves, in comparison with electric currents, which dissipate energy through ohmic losses.
This fact, besides its intrinsic interest from the fundamental physics point of view, makes magnonics a very active field
of research nowadays \cite{Pirro2021,Barman2021,Yu2021,Chumak2015}.
Indeed, several kinds of logical devices based on spin waves have been proposed, as magnonic logic gates \cite{Schneider2008},
magnonic logic circuits \cite{Khitun2010}, and a magnon transistor \cite{Chumak2014}.

To develop a technology based partly in spin waves it is necessary to have materials with adequate magnetic properties,
especially in what concerns the attenuation of spin waves. Ultralow magnetic damping is shown by some insulators, notably 
the yttrium iron garnet \cite{Yu2014,Hauser2016}, and 
has also been recently reported in thin films of a family of Heusler half-metals \cite{Guillemard2019}.
It is also necessary to have means to control and manipulate spin waves. This can be achieved in part by controlling the
magnetic textures on which spin waves propagate, either by manipulating them externally, producing graded magnetic textures
\cite{Davies2015a,Davies2015b,Dzyapko2016,Vogel2015,Vogel2018}, or by exploiting the inhomogeneous magnetic states
characteristic of chiral magnets, as skyrmion and one dimensional chiral soliton lattices. These states have the advantage
of appearing spontaneously and being controllable by external means like temperature or magnetic field
\cite{Bogdanov1994a,Muhlbauer2009,Yu2010,Laliena2017b,Laliena2018c,Togawa2012,Laliena2016b,Laliena2017a}.

One tool to control the spin waves is the scattering (reflection and transmission) at artificially created interfaces,
or at artificial magnetic patterns. This scattering induces interesting effects like Goos-H\"anchen displacements
\cite{Dadoenkova2012,Gruszecki2014,Gruszecki2017,Mailyan2017,Stigloher2018,Wang2019,Zhen2020}, the
Hartman effect \cite{Klos2018} and the Talbot effect \cite{Golebiewski2020}, which could be used to manipulate the
spin wave.

Spin waves are also scattered by magnetic solitons like domain walls \cite{Braun1994}, skyrmions
\cite{Schutte2014} or one dimensional chiral solitons \cite{Laliena2021}, producing effects that could also be useful to
control the spin waves. For instance, the scattering by a one dimensional soliton causes a lateral shift of the
propagation direction of the scattered waves analogous to the Goos-H\"anchen displacement \cite{Laliena2021}.
It has been proposed that the scattering by domain walls can be used for spin wave interferometry \cite{Hertel2004}
or as a spin wave valve \cite{Hamalainen2018}.
The scattering by solitons has the additional advantage that these kind of magnetic structures can be moved across
the material under the action of external influences like magnetic fields or electric currents
\cite{Schryer1974,Thiaville2005,Woo2016,Laliena2020,Osorio2021}.

In this paper we study the scattering of spin waves by a Bloch domain wall in an anisotropic ferromangnet.
It is known that such a domain wall is transparent to the spin waves since the reflection coefficient does
vanish.
This result is based on theoretical computations that either ignore the dipolar interaction or approximate
it by an effective local anisotropy \cite{Braun1994,Winter1961,Thiele1973}. Here we show that the domain wall does
actually reflect the spin waves if the dipolar interaction is properly taken into account. We obtain the reflected and transmitted
amplitudes treating the dipolar interaction as a perturbation and using the distorted wave Born approximation.
Due to the nature of the dipolar interaction this approximation is not straightforward, and it is necessary to
split the spin wave operator into an operator that can be included in the ``unperturbed'' operator plus another localized
operator, suitable to be treated in the Born approximation.
The reflection coefficient thus obtained is non zero, but it vanishes for normal incidence, what agrees with the
numerical simulations of Hertel et al. \cite{Hertel2004}, which take into account properly the dipolar interaction.


\section{The domain wall of an anisotropic ferromagnet}
\label{sec:model}

Let us consider a ferromagnet with uniaxial anisotropy of easy-axis type at a temperature sufficiently low, so
that the fluctuations of the modulus of the magnetization, $M_s$, can be neglected.
Then its magnetization is characterized by a unit vector field $\vn$.
We use a cartesian coordinate system with axes given by the
three orthonormal vectors $\vx$, $\vy$ and $\vz$ and coordinates $x$, $y$ and $z$ along these axes.
The points of space are represented by vectors like $\vrv$, with $x=\vx\vdot\vrv$, etc., and $r=|\vrv|$.
We will also use sometimes the notation $x_1=x$, $x_2=y$, and $x_3=z$, and $\vx_1=\vx$, $\vx_2=\vy$ and $\vx_3=\vz$,
and then $x_i=\vx_i\vdot\vrv$.
The magnet is oriented so that its anisotropy axis coincides with $\vz$.
The dynamics of the magnetization is derived from the energy functional $\mathcal{E}=\int d^3r w(\vrv)$ with
\be
w(\vrv) =
A\sum_{i=1}^3(\partial_{x_i}\vn)^2-K_u(\vz\vdot\vn)^2-\frac{\mu_0M_s^2}{2}\,\vn\vdot\vh_d, \label{eq:energy}
\ee
where the succesive terms in $w(\vrv)$ correspond to the
ferromagnetic exchange interaction, the anisotropy interaction, and the
dipolar interaction.
The constants $A>0$ and $K_u>0$ represent the strengths of the exchange and anisotropy interaction, respectively, and
$\mu_0$ is the vacuum permeability. The vector field $\vh_d$ is the dimensionless magnetostatic field, which is the solution
of the boundary value problem
\be
\nabla\cross \vh_d=0, \quad \nabla\vdot\vh_d=-\nabla\vdot\vn, \label{eq:maxwell}
\ee
in the whole space (interior and exterior to the magnet), with $\vh_d$ decaying sufficiently fast as
$r\to\infty$ as a condition.

The dynamics of the magnetization obeys the Landau-Lifschitz-Gilbert equation,
\be
\partial_t \vn = \gamma \vB_{\mathrm{eff}}\cross \vn + \alpha \vn\cross \partial_t\vn,
\ee
where $\gamma$ is the electron gyromagnetic factor, $\alpha$ is the Gilbert damping constant, and
$\vB_{\mathrm{eff}}$ is the effective field, given by the variational derivative (the first variation) of
the energy functional: $\vB_{\mathrm{eff}}=-(1/M_s)\delta\mathcal{E}/\delta\vn$. In the present case it is
\be
\vB_{\mathrm{eff}} = \frac{2A}{M_s}\Big(\nabla^2\vn + q_0^2 (\vz\vdot\vn)\vz + \epm q_0^2\,\vh_d\Big),
\ee
where $q_0=\sqrt{K_u/A}$ has the dimensions of inverse length and $\epm=\mu_0M_s^2/2K_u$ is dimensionless.
Notice that at a fixed time $\vh_d$ is a linear functional of $\vn$, given by the solution of (\ref{eq:maxwell}).
Since we are interested in the scattering of spin waves,
we neglect the damping term, assuming that the spin waves are able to propagate to long enough distances without
appreciable attenuation.

Let us consider a large magnet, which eventually will be infinite.
Let $L_x$, $L_y$, and $L_z$ be the system dimensions along the $\vx$, $\vy$ and $\vz$
directions, respectively, and let $L_z$ be much larger than $L_x$ and $L_y$.
In the limit $L_z\to\infty$ the ferromagnetic state with uniform magnetization along the $\vz$ direction is
an equilibrium state, since the magnetostatic field inside the magnet vanishes in this limit, and therefore
the energy functional attains its absolute minimum\footnote{Strictly speaking, we have to consider the energy
  density functional defined by $\mathcal{E}/L_xL_yL_z$.}.
After the $L_z\to\infty$ limit we take $L_x\to\infty$ and $L_y\to\infty$.
By symmetry, the uniform state with magnetization pointing along the $-\vz$ direction is another equilibrium
state.

This system has domain walls as metastable states. To see this, let us neglect first the dipolar interaction.
It is well known that the Euler-Lagrange equations of the functional (\ref{eq:energy}) with the dipolar
interaction term removed have the solution
\be
\vn_0(\vrv) = \sin\theta(x)\vy + \cos\theta(x)\vz, \label{eq:dw}
\ee
where $\theta(x) = 2\atan\big(\e^{q_0x}\big)$.
This state is a domain wall centered at $x=0$, which separates a domain with
$\vn(\vrv)\to\vz$ for $x\to -\infty$ from the opposite domain, with $\vn_0(\vrv)\to -\vz$, for $x\to +\infty$.
The magnetostatic field produced by the magnetization field (\ref{eq:dw}) vanishes in the infinite system, and
therefore (\ref{eq:dw}) is a solution of the Euler-Lagrange equations with dipolar interaction.
Moreover, the dipolar energy reaches its minimum (zero) at the domain wall state, which consequently
remains as a metastable state when the dipolar interaction is taken into account.

\section{Spin wave operator in presence of a domain wall}
\label{sec:swo}

Let us consider perturbations of the domain wall state, which in general can be described by two real
fields $\xi_1$ and $\xi_2$, so that
\be
\vn = \big(1+\xi_1^2+\xi_2^2\big)^{1/2}\vn_0 + \xi_1\ve_1 + \xi_2\ve_2,
\ee
where $\{\ve_1,\ve_2,\vn_0\}$ is a right-handed orthonormal triad. Notice that $\ve_1$ and $\ve_2$ depend
on $\vrv$, since $\vn_0$ does. We take
\be
\ve_1(\vrv) = \vx, \quad \ve_2(\vrv) = \cos\theta(x)\vy - \sin\theta(x)\vz.
\ee

We consider local perturbations, $\delta\vn=\xi_1\ve_1+\xi_2\ve_2$, whose absolute value decreases to zero
rapidly enough as $r\to\infty$. These local perturbations propagate through the magnet as spin waves.
Their dynamics are governed by the linearized Landau-Lifschitz-Gilbert equation, which,
neglecting the damping term, has the form
\be
\partial_t\delta\vn = \gamma \vB_{\mathrm{eff}}^{(0)}\cross\delta\vn + \gamma \delta\vB_{\mathrm{eff}}\cross \vn_0,
\label{eq:llgp}
\ee
where $\vB_{\mathrm{eff}}^{(0)}$ is the effective field corresponding to the
metastable state $\vn_0$
\be
\vB_{\mathrm{eff}}^{(0)}=\frac{2A}{M_s}q_0^2\cos(2\theta)\vn_0,
\ee
and $\delta\vB_{\mathrm{eff}}$ is the effective field to first order in the perturbation $\delta\vn$,
\be
\delta\vB_{\mathrm{eff}} = \frac{2A}{M_s}\Big(\nabla^2\delta\vn + q_0^2 (\vz\vdot\delta\vn)\vz
+ \epm q_0^2\,\delta\vh_d\Big),
\ee
with $\delta\vh_d$ being the magnetostatic field created by the perturbation, which is the solution of
\be
\nabla\cross \delta\vh_d=0,\quad \nabla\vdot\delta\vh_d=-\nabla\vdot\delta\vn. \label{eq:maxwell2}
\ee
Projecting equation (\ref{eq:llgp}) onto $\ve_1$ and $\ve_2$ we obtain the equations for the dynamics of
$\xi_1$ and $\xi_2$:
\bea
  &&\partial_t \xi_1 = -\pto\xi_2 + \omega_0\,\epm\,(\delta\vh_d\cross\vn_0)\vdot\ve_1, \\[4pt]
  &&\partial_t \xi_2 = \pto\xi_1 + \omega_0\,\epm\,(\delta\vh_d\cross\vn_0)\vdot\ve_2,
\eea
where $\omega_0=2\gamma Aq_0^2/M_s$ and $\pto$ is the Schr\"odinger operator
\be
\pto = -\frac{\omega_0}{q_0^2}\nabla^2 + \omega_0 - 2\omega_0\sech^2(q_0x).
\ee
The dipolar field determined by equations (\ref{eq:maxwell2}) is linear in $\xi_1$ and $\xi_2$ and thus we have
\bea
&&\big[\delta\vh_d(\vrv)\cross\vn_0(x)\big]\vdot\ve_1(x) = \big(D_{11}\xi_1\big)(\vrv) + \big(D_{12}\xi_2\big)(\vrv),
\label{eq:Ddef1} \\[4pt]
&&\big[\delta\vh_d(\vrv)\cross\vn_0(x)\big]\vdot\ve_2(x) = \big(D_{21}\xi_1\big)(\vrv) + \big(D_{22}\xi_2\big)(\vrv),
\label{eq:Ddef2}
\eea
where the $D_{\alpha\beta}$ are linear operators which will be determined in the next section. Thus, defining $\xi$ as
the two-component column vector $\xi=(\xi_1,\xi_2)^T$, the spin wave equation can be written as
\be
\partial_t\xi = \Omega\xi,
\ee
where $\Omega=\Omega_0 + \epm \omega_0 D$ is a linear operator with
\be
\Omega_0 = \begin{pmatrix} 0 & -\pto\\ \pto & 0 \end{pmatrix}, \quad
D = \begin{pmatrix} D_{11} & D_{12} \\ D_{21} & D_{22} \end{pmatrix}.
\ee

If the dipolar interaction is neglected, or if it is approximated by an effective interaction included in $K_u$,
the dynamics of the spin waves is given by $\Omega_0$.
This operator has been studied since long ago by a number of researchers
(see references
\onlinecite{Winter1961,Thiele1973,Braun1994,Hertel2004,Bayer2005,Kishine2011,Borys2016,Whitehead2017}).
Let us recall its spectral properties, which are needed in the following.
Let $\psi$ be an eigenfunction of $\pto$, with eigenvalue $\nu\geq 0$
(since the spectrum of $\pto$ is non negative), so that $\pto\psi=\nu\psi$. Then the two states
\be
 \frac{1}{\sqrt{2}}\begin{pmatrix} 1 \\ -\iu \end{pmatrix} \psi, \quad
\frac{1}{\sqrt{2}}\begin{pmatrix} 1 \\ \iu \end{pmatrix} \psi, \label{eq:eigenstates}
\ee
are eigenstates of $\Omega_0$ with eigenvalues $+\iu\nu$ and $-\iu\nu$, respectively.
Hence, the spectral properties of $\Omega_0$ are fully determined by those of $\pto$.

To obtain the spectrum of $\pto$ we perform a Fourier transform in the variables $y$ and $z$,
\be
\tpsi(x,\vkt) = \int d^2\rt \e^{-\iu\vkt\vdot\vrvt}\psi(x,\vrvt),
\ee
where $\vkt=k_y\vy+k_z\vz$ and $\vrvt=y\vy+z\vz$, and the spectral equation for $\pto$ becomes
\be
\frac{\omega_0}{q_0^2}\Big(-\frac{d^2}{dx^2} + \kt^2+ q_0^2 - 2q_0^2\sech^2(q_0x)\Big)\tpsi(x,\vkt)
= \nu\,\tpsi(x,\vkt).
\ee
This is a one dimensional time independent Schr\"odinger equation with potential $-2q_0^2\sech^2(q_0x)$,
which is exactly solvable \cite{Drazin1989}.
Its spectrum consists of one bound state with eigenvalue $\nu_B=\omega_0\kt^2/q_0^2$ and eigenfunction
\be
\phi_B(x) = \frac{q_0}{\sqrt{2}}\sech(q_0x),
\ee
and a continuum spectrum above a gap, given by $\omega_G=\nu_B+\omega_0$.
The continuum spectrum is parametrized by a real number (wave number) $k_x$ as
\be
\nu(\vk)=\omega_0\frac{k_x^2}{q_0^2}+\omega_G,
\ee
with $\vk=k_x\vx+\vkt$, and has the eigenfunctions
\be
\phi_{k_x}(x) = \frac{1}{\sqrt{q_0^2+k_x^2}}\e^{\iu k_xx}\big(q_0\tanh(q_0x)-\iu k_x\big).
\ee
The eigenfunctions satisfy the normalization condition
\begin{gather}
\int_{-\infty}^\infty \phi_B^2(x) dx = 1, \\
\int_{-\infty}^\infty \phi_{k_x}(x)^*\phi_{k_x^\prime}(x)dx = \delta(k_x-k_x^\prime), 
\end{gather}
and the closure relation
\be
\phi_B(x)\phi_B(x^\prime) + \int_{-\infty}^\infty \frac{dk_x}{2\pi} \phi_{k_x}(x)\phi_{k_x}^*(x^\prime) 
= \delta(x-x^\prime). \label{eq:closure}
\ee

The eigenstates of $\Omega_0$ are obtained by substituting $\psi$ in (\ref{eq:eigenstates}) by
$\e^{\iu\vkt\vdot\vrvt}\phi_B(x)$ or by $\e^{\iu\vkt\vdot\vrvt}\phi_{k_x}(x)$. The closure relation (\ref{eq:closure})
ensures that $\Omega_0$ has the spectral representation
\be
\Omega_0(\vrv,\vrv^\prime) = \int \frac{d^2\kt}{(2\pi)^2} \e^{\iu\vkt\vdot(\vrvt-\vrvt^\prime)}\tilde{\Omega}_0(\vkt,x,x^\prime),
\ee
where
\begin{widetext}
\be
\tilde{\Omega}_0(\vkt,x,x^\prime) = \begin{pmatrix} 0 & -\nu_B \\ \nu_B & 0 \end{pmatrix} \phi_B(x)\phi_B(x^\prime)
+ \int_{-\infty}^\infty \frac{dk_x}{2\pi} \begin{pmatrix} 0 & -\nu(\vk) \\ \nu(\vk) & 0 \end{pmatrix}
\phi_{k_x}(x)\phi_{k_x}^*(x^\prime).
\ee
\end{widetext}
From now on we will not show explicitely the $\vkt$ dependence of $\tilde{\Omega}_0$, which has to be 
understood.

The spin wave spectrum contains two states bound to the domain wall, sometimes called Winter modes \cite{Winter1961},
whose spatial distribution is described by the wave function $\phi_B(x)$, which decays exponentially for
$|x|\to\infty$. These modes are very interesting since they only propagate on the domain wall plane, so that they
might be used as a wave guide for spin waves \cite{GarciaSanchez2015}. Spin wave propagation 
bound to the domain wall has been experimentally observed by Wagner et al. \cite{Wagner2016}.

In this paper, however, we focus on the scattering of unbounded spin waves by the domain wall.
For that we will need the asymptotic behavior of $\phi_{k_x}(x)$ as $x\to\pm\infty$, which is given
by
\be
\phi_{k_x}(x) \sim -\iu\,\e^{\pm\iu\delta_0}\e^{\iu k_xx},\quad x\to\pm\infty,
\ee
where
\be
\delta_0 = \pi/2-\atan(k_x/q_0).
\ee

It is well known that the $2q_0^2\sech^2(q_0x)$ potential is reflectionless \cite{Drazin1989}, and this quality
is inherited by the $\Omega_0$ operator. Therefore the domain wall does not reflect the spin waves if the dipolar
interaction is neglected, or if it is approximated by an effective magnetic anisotropy
\cite{Braun1994,Winter1961,Thiele1973}.

\section{The contribution of the dipolar interaction \label{sec:df}}

Let us analyze the form of the $D$ operator, which gives the contribution of the dipolar interaction to the
spin wave operator.

Since we consider local perturbations which vanish sufficiently rapid as $r\to\infty$, the solution of
equations (\ref{eq:maxwell2}) is
\be
\delta\vh_d(\vrv) =
-\frac{1}{4\pi}\int d^3r^\prime\,\frac{\vrv-\vrv^\prime}{|\vrv-\vrv^\prime|^3}\,\nabla\vdot\delta\vn(\vrv^\prime).
\label{eq:df}
\ee
Combining this expression with equations (\ref{eq:Ddef1}) and (\ref{eq:Ddef2}) we obtain the form of the $D$
operator.
Noticing that $\ve_1$ and $\ve_2$ are independent of $y$ and $z$, we perform the Fourier expansion in the 
variables $y$ and $z$ (recall that $\vrvt=y\vx+z\vz$ and $\vkt=k_y\vx+k_z\vz$):
\be
\xi_\alpha(x,\vrvt)=\int \frac{d^2\kt}{(2\pi)^2} \e^{\iu\vkt\vdot\vrvt}\txi_\alpha(x,\vkt),\;\;\alpha=1,2.
\label{eq:xif}
\ee
In this way we get
\be
\big(D_{\alpha\beta}\xi_\beta\big)(x,\vrvt) = \int \frac{d^2\kt}{(2\pi)^2} \e^{\iu\vkt\vdot\vrvt}
\big(\tilde{D}_{\alpha\beta}\txi_\beta\big)(x,\vkt),
\ee
where no summation in $\beta$ is is to be understood and
\be
\big(\tilde{D}_{\alpha\beta}\txio\big)(x,\vkt) =
\int_{-\infty}^\infty \!\!dx^\prime \tilde{D}_{\alpha\beta}(x,x^\prime,\vkt) \txi_\beta(x^\prime,\vkt),
\ee
where the kernels are given by
\bea 
\tilde{D}_{11}(x,x^\prime,\vkt) &=& -\iu\Fx(x,\vkt)\sigma(x-x^\prime)\rho(x-x^\prime), \;\;
\\[2pt]
\tilde{D}_{12}(x,x^\prime,\vkt) &=& -\Fx(x,\vkt) \rho(x-x^\prime) \Fx(x^\prime,\vkt)
\\[2pt]
\tilde{D}_{21}(x,x^\prime,\vkt) &=& \delta(x-x^\prime) - \rho(x-x^\prime),
\\[2pt]
\tilde{D}_{22}(x,x^\prime,\vkt) &=& \iu \sigma(x-x^\prime) \rho(x-x^\prime) \Fx(x^\prime,\vkt).
\eea
In these expressions we introduce the functions 
\be
\sigma(x)=x/|x|, \quad \rho(x) = \frac{\kt}{2}\e^{-\kt|x|}.
\ee
and $F(x,\vkt)$, which is the projection of $\vkt/\kt$ onto $\ve_2(x)$:
\be
\Fx(x,\vkt) = \frac{k_y}{\kt}\cos\theta(x) - \frac{k_z}{\kt}\sin\theta(x). \label{eq:F}
\ee
Notice that $F(x,\vkt)\to\mp k_y/\kt$ as $x\to\pm\infty$.
Some details on the derivations of the operators $\tilde{D}_{\alpha\beta}$ are given in appendix A.

For fixed $\vkt$ the operator $\tilde{D}$ is not invariant under reflection about the domain wall center, $x=0$.
This is due to the fact that the the equilibium state $\vn_0(x)$ is not invariant under reflection with
respect to the $\vy\vz$ plane (not even the ferromagnetic state $\vn_0(x)=\vz$ is invariant, since $\vn$ is an axial vector).
However, $\vn_0(x)$ is invariant under the composition of a reflection with respect
to $\vy\vz$ plane and a reflection with respect to the $\vx\vy$ plane. This means that $\tilde{D}$ is invariant
under the transformation $x\to-x$ and $k_z\to-k_z$, keeping $k_y$ unchanged, as can be easily checked.

The operator $D$ contributes to the dynamics of the asymptotic spin wave states, since
$\big(\tilde{D}_{\alpha\beta}\txi_\alpha\big)(x,\vkt)$ does not vanish as $|x|\to\infty$.
It is clear that this has to be so since the dipolar interaction affects also to the perturbations of
the ferromagnetic states.
To study the scattering we have to separate from $\tilde{D}_{\alpha\beta}$ the part that survives as $|x|\to\infty$.
Let us introduce the asymptotic operators $\Dpm_{\alpha\beta}$ so that
\be
\tilde{D}_{\alpha\beta}\txi_\beta(x,\vkt) \sim \Dpm_{\alpha\beta}\txi_\beta(x,\vkt)
\ee
for $x\to\pm\infty$.
Taking into account the asymptotic behaviour of $F(x,\vkt)$ as $x\to\pm\infty$ we have
\bea
\Dpm_{11}(x,x^\prime,\vkt) &=& \mp\iu\frac{k_y}{\kt}\sigma(x-x^\prime)\rho(x-x^\prime),
\\[2pt]
\Dpm_{12}(x,x^\prime,\vkt) &=& -\frac{k_y^2}{\kt^2}\rho(x-x^\prime),
\\[2pt]
\Dpm_{21}(x,x^\prime,\vkt) &=& \delta(x-x^\prime) - \rho(x-x^\prime),
\\[2pt]
\Dpm_{22}(x,x^\prime,\vkt) &=& \pm\iu\frac{k_y}{\kt}\sigma(x-x^\prime)\rho(x-x^\prime).
\eea

The two asymptotic operators are different due obviously to the fact that spin waves propagate on 
ferromagnetic domains with opposite magnetization if $x\to -\infty$ and $x\to +\infty$.
To avoid the complications of scattering with two
different asymptotic operators we consider $k_y=0$. In this case $F(x,\vkt)$ tends to zero exponentially
as $|x|\to\infty$ and therefore the only nonvanishig
asymptotic operators are $\Dm_{21}=\Dp_{21}$, and therefore we have a single asymptotic operator
for $|x|\to\infty$.

The simplicity of the asymptotic $D$ operator in the case $k_y=0$ (only $\Dpm_{21}$ is non zero)
can be easily understood: the perturbations for $x\to\pm\infty$ are $\delta\vn \sim \xi_1\vx \mp \xi_2\vy$
and therefore the source of the dipolar field is
\be
\nabla\vdot\delta\vn \sim \partial_x\xi_1 \mp \partial_y\xi_2.
\ee
Since $\partial_y\xi_2=0$ if $k_y=0$, we have that in this case the dipolar interaction depends only on
$\partial_x\xi_1$ if $x\to\pm\infty$. Hence the asymptotic $D$ operator acts only on $\xi_1$ and is the same
for $x\to\pm\infty$.

The asymptotic operators are translationally invariant and their kernels
have a Fourier representation which for $k_y=0$ is given by 
\be
\Dpm(x-x^\prime) = Z \int_{-\infty}^\infty\frac{dk_x}{2\pi}\,\frac{k_ x^2}{k_x^2+\kt^2} \e^{\iu k_x(x-x^\prime)},
\ee
where
\be
Z = \begin{pmatrix} 0 & 0 \\ 1 & 0 \end{pmatrix}. \label{eq:Das}
\ee
Notice that we use the same symbol for the operators $\Dpm$ and their integral kernels.

\section{The scattering problem}
\label{sec:sp}

We address the scattering problem perturbatively, taking advantage of the exact solvability of the problem in
absence of the dipolar interaction and treating this as a perturbation. To this end we have to separate the
spin wave operator into an operator which is to be treated exacly (it has to contain $\tilde{\Omega}_0$) and has the correct
asymptotic behaviour, plus a localized perturbation which does not contribute to the dynamics of the asymptotic states.
Localization means that the operator is given by an
integral kernel $\Theta(x,x^\prime)$ so that the integral $\int |\Theta(x,x^\prime)| dx^\prime$ decays exponentially
to zero as $|x|\to\infty$. A particular case of this is a potential that decays rapidly enough with the distance.
However, the perturbation in the present case does not have the form of a potential.

\subsection{Split of the spin wave operator into an ``unperturbed'' operator plus a perturbation}

The perturbation cannot be $\epm \omega_0 \tilde{D}$ since this is not a localized operator.
In the case $k_y=0$, we can separate from $\tilde{D}$ its asymptotic part, $\Dm$,
and $\tilde{D}-\tilde{D}^{(-)}$ is localized.
The problem with this natural identification of the perturbation is that we do not have the exact
spectrum of $\Oo+\Dm$. To overcome this difficulty we split $\Dm$ as
$\Dm=\Ds+\Delta$, where these two new operators are given by the integral kernels
\be
\Ds(x,x^\prime) = Z \int\frac{dk_x}{2\pi}\,\frac{k_ x^2}{k_x^2+\kt^2} \phi_{k_x}(x)\phi_{k_x}^*(x^\prime)
\ee
and
\be
\Delta(x,x^\prime) = Z\!\int\frac{dk_x}{2\pi}\,\frac{k_ x^2}{k_x^2+\kt^2}
\Big(\e^{\iu k_x(x-x^\prime)}-\phi_{k_x}(x)\phi_{k_x}^*(x^\prime)\Big). \label{eq:Delta1}
\ee
The sum of these two operatos give $\Dm$, as can be seen from equation (\ref{eq:Das}).
The key points are: i) $\Ds$ has the asymptotic behaviour of $\Dm$ 
and $\Os = \Oo + \epm \omega_0 \Ds$ is an ``unperturbed'' operator that can be treated exactly
and has the correct the asymptotic behaviour; and ii) that, as we show below, $\Delta$ is a localized operator.
The reason for this is that the spectral projector $\phi_{k_x}(x)\phi_{k_x}^*(x^\prime)$ tends asymptotically to
$\exp\big(\iu k_x (x-x^\prime)\big)$, the difference between these two functions being
a function exponentially decaying with $|x|$.

Summarizing, we have split the spin wave operator into an ``unperturbed'' term $\Os$ and a localized perturbation
$V$ as
\be
\tilde{\Omega} = \Os + \epm \omega_0 V, \label{eq:split}
\ee
where $V=\tilde{D}-\Dm + \Delta$. Equation (\ref{eq:split}) is the key point of this work.

\subsection{The $\Os$ operator}

To study the scattering we need the asymptotic states, which are given by the eigenstates of $\Os$.
The explicit form of the integral kernel of $\Os$ is given by
\be
\Os(x,x^\prime) = \Os^{(\mathrm{b})}(x,x^\prime) + \Os^{(\mathrm{s})}(x,x^\prime), \label{eq:osk}
\ee
with
\be
\Os^{(\mathrm{b})}(x,x^\prime) =
\begin{pmatrix} 0 & -\nu_B \\ \nu_B & 0 \end{pmatrix} \phi_B(x)\phi_B(x^\prime), \label{eq:oskb}
\ee
and
\be
\Os^{(\mathrm{s})}(x,x^\prime) = \int\frac{dk_x}{2\pi}
\begin{pmatrix}
  0 & -\omega_2(\vk) \\ \omega_1(\vk) & 0
\end{pmatrix}
\phi_{k_x}(x)\phi_{k_x}^*(x^\prime), 
\label{eq:osks}
\ee
where we define
\be
\omega_1(\vk) = \nu(\vk)+ \epm \frac{\omega_0 k_ x^2}{k_x^2+\kt^2}, \quad \omega_2(\vk) = \nu(\vk).
\ee

The spectrum of $\Os$ consists of the two bound states of $\Oo$ and a continuum of states, with spectrum on the
imaginary axis parametrized by $k_x$ as $\pm\iu\sqrt{\omega_1\omega_2}$, and with eigenstates given by
\be
\phi_{k_x}(x)\,\up, \quad \phi_{k_x}(x)\,\um,
\ee
where the labels $p$ and $m$ correspond to $+\iu\sqrt{\omega_1\omega_2}$ and $-\iu\sqrt{\omega_1\omega_2}$,
respectively. In the above expressions we introduced the two component vectors
\be
\xi_m = \frac{1}{\big(\omega_1+\omega_2\big)^{1/2}}
  \begin{pmatrix} \sqrt{\omega_2} \\ \iu\sqrt{\omega_1} \end{pmatrix},
\ee
and $\xi_p=\xi_m^*$.
For fixed $\kt$ each eigenvalue is doubly degenerate, the degeneracy corresponding to the two opposite values
of $k_x$, since $\omega_1$ and $\omega_2$ are even functions of $k_x$.

\subsection{The $\Delta$ operator}

Let us write $\Delta(x^\prime,x^{\prime\prime})=d(x^\prime,x^{\prime\prime})Z$, so that $d(x^\prime,x^{\prime\prime})$ is
the integral entering the left hand side of equation (\ref{eq:Delta1}). Taking into account the form of
$\phi_{k_x}(x)$ we get
\be
d(x^\prime,x^{\prime\prime}) = \int \frac{dk_x}{2\pi}\e^{\iu k_x(x^\prime-x^{\prime\prime})}
\frac{q_0^2\,k_x^2\,g(x^\prime,x^{\prime\prime},k_x)}{(k_x^2+\kt^2)(k_x^2+q_0^2)},
\ee
where
\be
\begin{split}
g(x^\prime,x^{\prime\prime},k_x) = 1-\tanh(q_0x^\prime) \tanh(q_0x^{\prime\prime}) \\
+ \,\iu\,\frac{k_x}{q_0}\big(\tanh(q_0x^{\prime\prime}) - \tanh(q_0x^\prime)\big). \hspace*{0.7cm}
\end{split}
\ee
If $x^\prime\neq x^{\prime\prime}$ the integrand behave for large $|k_x|$ as
$\exp\big(\iu k_x(x^\prime-x^{\prime\prime})\big)/k_x$, which is integrable, while it behaves as $1/k_x^2$
if $x^\prime=x^{\prime\prime}$, which is also integrable. The integral can be evaluated by the method of residues,
closing the integration contour
on the upper half complex plane if $x^\prime-x^{\prime\prime}>0$ or on the lower half complex plane if 
$x^\prime-x^{\prime\prime}<0$. We obtain
\be
d(x^\prime,x^{\prime\prime}) = \frac{q_0^2}{\kt^2-q_0^2}
\Big(d_0(x^\prime,x^{\prime\prime},\kt)-d_0(x^\prime,x^{\prime\prime},q_0)\Big),
\ee
where
\be
\begin{split}
d_0(x^\prime,x^{\prime\prime},k) = \Big(1+\tanh(q_0x^\prime) \tanh (q_0x^{\prime\prime}) \hspace*{1.0cm} \\
+\frac{k}{q_0}\big|\tanh (q_0x^\prime) - \tanh (q_0x^{\prime\prime})\big|\Big)
\frac{k}{2}\e^{-k|x^\prime-x^{\prime\prime}|}.
\end{split}
\ee
It is easily checked that the kernel $d(x^\prime,x^{\prime\prime})$ is continuous at $\kt=q_0$.
We see that, as expected, $\Delta$ is a localized operator.

\subsection{The Lippmann-Schwinger equation}

The spectral equation for $\tilde{\Omega}$ has the form
\be
\big(\Os + \epm \omega_0 V\big)\xi = -\iu\omega\xi.
\ee
We henceforth consider on $\omega>0$ and $k_x>0$, where $k_x$ is related to $\omega$ by
$\sqrt{\omega_1\omega_2}=\omega$.
Since $V$ is a localized operator, the solutions of the above equation behave asymptotically as
eigenstates of $\Os$, that is
\be
\xi_{k_x} \sim \big(\alpha_\pm\,\e^{\iu k_xx} + \beta_\pm\,\e^{-\iu k_xx}\big)\um
\ee
for $x\to\pm\infty$, taking into account the asymptotic behavior of $\phi_{k_x}(x)$.
The solution appropriate for scattering requires $\beta_+=0$ (no wave incoming from
$+\infty$), and in this case $\beta_-/\alpha_-$ and $\alpha_+/\alpha_-$ are the reflected and transmitted amplitudes,
respectively.

The condition $\beta_+=0$ is satisfied if the eigenstate $\xi_{k_x}^+$ is chosen as the solution of the
Lippmann-Schwinger equation
\be
\begin{split}
\hspace*{-0.0cm} \xi_{k_x}^+(x) = \phi_{k_x}(x)\,\um + \int_{-\infty}^\infty \!\!dx^\prime G^+(x,x^\prime,-\iu\omega+\mu) \times  \\
\int_{-\infty}^\infty \!\!dx^{\prime\prime}\epm \omega_0 V(x^\prime,x^{\prime\prime})\,\xi_{k_x}^+(x^{\prime\prime}), \hspace*{1.2cm}
\end{split}
\label{eq:lse}
\ee
with $\mu\to 0^+$.
The Green's function $G^+$ is the integral kernel of the resolvent operator
$(-\iu\omega+\mu-\Os)^{-1}$, and satisfies the asymptotic condition
\be
\lim_{\mu\to 0^+}G^+(x,x^\prime;-\iu\omega+\mu) \sim \e^{\iu k_x x} Q(x^\prime)
\ee
for $x\to +\infty$, where $Q(x^\prime)$ is a $2\times 2$ matrix independent of $x$.
The positive sign of $\mu$ ensures that this condition holds, as will be seen below.

\section{The Green's function}
\label{sec:gf}

The scattering parameters are obtained from the asymptotic behavior of $\xi_{k_x}^+$ as
$x\to\pm\infty$. Therefore, to calculate them we need the asymptotic behavior of the Green's function.

Using the spectral representation (\ref{eq:osk}) we obtain
\begin{widetext}
\be
\begin{split}
G^+(x,x^\prime,-\iu\omega+\mu) =
\frac{1}{(-\iu\omega+\mu)^2+\nu_B^2}
\begin{pmatrix}
-\iu\omega+\mu & -\nu_B \\ \nu_B & -\iu\omega+\mu \\
\end{pmatrix} \phi_B(x)\phi_B(x^\prime) 
\\[4pt]
+ \int_{-\infty}^\infty \frac{dk_x^\prime}{2\pi}\frac{1}{(-\iu\omega+\mu)^2+\omega_1\omega_2}
\begin{pmatrix}
-\iu\omega+\mu & -\omega_2 \\ \omega_1 & -\iu\omega+\mu \\
\end{pmatrix} \phi_{k_x^\prime}(x)\phi_{k_x^\prime}^*(x^\prime), \hspace*{0.4cm}
\end{split}
\label{eq:gfs}
\ee
\end{widetext}
where it is understood that $\omega_1$ and $\omega_2$ depend on $k_x^\prime$.
As we said, we reserve the symbol $k_x$ for the solutions of $\sqrt{\omega_1\omega_2}=\omega$.

\begin{figure}[t!]
\centering
\includegraphics[width=0.7\linewidth,angle=0]{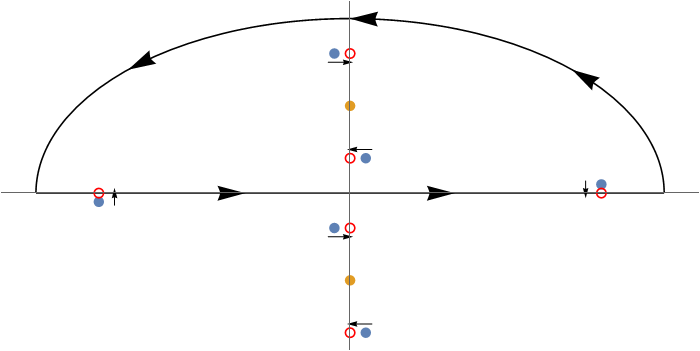}
\caption{Pole structure of the integrand of the right hand side of equation (\ref{eq:gfs}).
\label{fig:poles}}
\end{figure}

The part of the Green's function due to the bound states does not contribute to the asymptotic
behavior, and it can be safely ignored since we take $\omega$ above the gap ($\omega>\nu_B+\omega_0$).

Thus, we have to evaluate the integral of the right hand side of equation (\ref{eq:gfs}) for
$x\to\pm\infty$. The integrand is a meromorphic function
of $k_x^\prime$ that
decays exponentially to zero as $|k_x^\prime|\to\infty$ on the upper half complex plane if $x>x^\prime$, and
on the lower half plane if $x<x^\prime$, due to the form of $\phi_{k_x^\prime}(x)$.
Therefore the integral can be evaluated by the method of residues,
choosing an integration contour as in figure \ref{fig:poles} for $x>x^\prime$.

The generic pole structure of the integrand, which is analyzed with some detail in appendix B, is displayed
in figure \ref{fig:poles}. 
There are two poles coming from $\phi_{k_x^\prime}(x)\phi_{k_x^\prime}^*(x^\prime)$, located on the imaginary axis
at $\pm\iu q_0$ (yellow points). In addition, there are six more poles (blue points),
three of them on the upper half plane and another three on the lower half plane (see figure \ref{fig:poles}).
As $\mu\to0^+$ two of these six poles attain the real axis, at the solutions $k_x^\prime=\pm k_x$ 
of the equation $\sqrt{\omega_1\omega_2} = \omega$ (see appendix B)
The negative pole $-k_x$ is reached from the lower half plane and the positive pole $k_x$ from the upper half
plane. All the other poles remain separated from the real axis as $\mu\to0^+$
(see the red circles in Figure \ref{fig:poles}).

Consider the case $x>x^\prime$.
For $x-x^\prime\to\infty$ the contribution of poles which do not attain the real axis as $\mu\to 0^+$ is
exponentially small and do not contribute
to the asymptotic behaviour, which is given only by the $k_x$ pole. Its residue can be readily computed and 
gives the asymptotic part, as $x\to\infty$, keeping $x^\prime$ fixed, of the Green's function
\be
G_{\mathrm{as}}^+(x,x^\prime,-\iu\omega) = -\frac{\iu}{\vgx}
\e^{\iu\delta_0}\e^{\iu k_x x}\phi_{k_x}^*(x^\prime)  \Pm, \label{eq:gfap}
\ee
where $\vgx=\partial\omega/\partial k_x$ is the group velocity, and
\be
\Pm=\frac{1}{2}\begin{pmatrix}
1 & -\iu\sqrt{\omega_2/\omega_1} \\ \iu\sqrt{\omega_1/\omega_2} & 1
\end{pmatrix}
\ee
is the projector along $\up$ onto $\um$:
\be
\Pm\,\up = 0, \quad \Pm\,\um = \um.
\ee
One has to bear in mind that in equation (\ref{eq:gfap}) $\omega_1$ and $\omega_2$ depend on $k_x$ and that
$\omega$ and $k_x$ are related by the equation $\omega_1\omega_2=\omega^2$ (the dispersion relation) which
then determines the group velocity.

For $x-x^\prime<0$ we have to close the integral contour on the lower half plane and again only the pole attaining the
real axis (this time at $k_x^\prime=-k_x$) as $\mu\to0^+$ contributes to the asymptotic behavior $x-x^\prime\to-\infty$.
The asymptotic Green's function as $x\to-\infty$ with $x^\prime$ fixed is given by
\be
G_{\mathrm{as}}^+(x,x^\prime,-\iu\omega) = \frac{\iu}{\vgx}
\e^{\iu\delta_0}\e^{-\iu k_x x}\phi_{k_x}(x^\prime)\Pm. \label{eq:gfan}
\ee

\section{Distorted wave Born approximation}
\label{sec:ba}

We get an approximation to $\xi_{k_x}^+(x)$ by using the first (distorted wave) Born approximation to solve the Lippmann-Schwinger
equation, substituting on its right-hand-side $\xi_{k_x}(x^{\prime\prime})$ by $\phi_{k_x}(x^{\prime\prime})\,\um$:
\be
\begin{split}
\hspace*{-0.1cm}
\xi_{k_x}^+(x) = \phi_{k_x}(x)\,\um + \int_{-\infty}^\infty dx^\prime G^+(x,x^\prime,-\iu\omega+\mu) \times \\
\int_{-\infty}^\infty dx^{\prime\prime}\epm \omega_0 V(x^\prime,x^{\prime\prime})\, \phi_{k_x}(x^{\prime\prime})\,\um.
\hspace*{0.8cm}
\end{split}
\label{eq:born}
\ee
We expect the Born approximation will be good if $\epm q_0/k_x$ is small enough, since the correction to the wave function
introduced by the perturbation considered here is of this order. It is well known that in one dimensional problems 
the Born approximation cannot be used in the vicinity of the gap frequency (small $k_x$), since the Green's function diverges
for $k_x\to 0$ (see reference \onlinecite{LandauQM}).

The scattering properties (reflection and transmission amplitudes) are obtained in the Born approximation from
the explicit expression for $\xi_{k_x}^+(x)$ given by equation (\ref{eq:born}).

\subsection{The $x\to\infty$ asymptotics}

For $x\to\infty$ we can substitute the Green's function by the corresponding asymptotic Green's function,
given by equation (\ref{eq:gfap}). We can neglect the contribution to the integral in $dx^\prime$ of the
region in which $x^\prime$ is of the order of, or larger than, $x$, since
\be
\int_{-\infty}^\infty dx^{\prime\prime}\epm \omega_0 V(x^\prime,x^{\prime\prime})\, \phi_{k_x}(x^{\prime\prime})
\ee
tends to zero exponentially as $x^\prime\to\infty$. This is due to the fact that the perturbation $V$ is a localized operator.
Using the asymptotic form of $\phi_{k_x}(x)$ and $G_{\mathrm{as}}^+$ given by equation (\ref{eq:gfap}),
we get for $x\to\infty$
\be
\xi_{k_x}^+(x) \sim -\iu\e^{\iu\delta_0}\e^{\iu k_x x} \um - \frac{\iu\epm \omega_0}{\vgx}\e^{\iu\delta_0}\e^{\iu k_x x}\Pm
\mathcal{T} \um,
\ee
where the $2\times 2$ matrix $\mathcal{T}$ depends only on $k_x$ and $k_z$ and is given by
\be
\mathcal{T} = \int_{-\infty}^\infty dx^\prime \phi_{k_x}^*(x^\prime)
\int_{-\infty}^\infty dx^{\prime\prime}V(x^\prime,x^{\prime\prime}) \phi_{k_x}(x^{\prime\prime}).
\ee
Taking into account the form of $V(x^\prime,x^{\prime\prime})$, the matrix elements $t_{ij}$ of $\mathcal{T}$ are
given by the integrals
\be
t_{ij} = \int dx^\prime \int dx^{\prime\prime} \phi_{k_x}^*(x^\prime) f_{ij}(x^\prime,x^{\prime\prime},k_z)\phi_{k_x}(x^{\prime\prime}),
\label{eq:tij}
\ee
where 
\bea
&&f_{11} = \iu\,\sigma(k_z) \sin\theta(x^\prime) \sigma(x^\prime-x^{\prime\prime})\rho(x^\prime-x^{\prime\prime}),
\label{eq:t11} \\[2pt]
&&f_{12} = -\sin\theta(x^\prime)\rho(x^\prime-x^{\prime\prime})\sin\theta(x^{\prime\prime}),
\label{eq:t12} \\[2pt]
&&f_{21} = d(x^\prime-x^{\prime\prime}), \label{eq:t21} \\[2pt]
&&f_{22} = -\iu\,\sigma(k_z) \sigma(x^\prime-x^{\prime\prime})\rho(x^\prime-x^{\prime\prime})
\sin\theta(x^{\prime\prime}). \hspace*{0.5cm}
\label{eq:t22}
\eea

Since $\Pm$ projects onto $\um$, we have
$\Pm\mathcal{T}\um=\chi_t\um$, where $\chi_t$ is a complex number that can be computed
in terms of the $t_{ij}$:
\be
\chi_t = \frac{\iu}{2}\big(\sqrt{\omega_1/\omega_2}\,t_{12}-\sqrt{\omega_2/\omega_1}\,t_{21}\big).
\ee
In deriving the above expression we used the fact that, by symmetry, $t_{11}+t_{22}=0$.
Furthermore, the integrals that define $t_{12}$ and $t_{21}$ can be evaluated explicitely in terms
of the derivative of the digamma function. The explicit expressions are given in appendix C.

Summarizing, we have obtained that for $x\to\infty$
\be
\xi_{k_x}^+(x) \sim -\iu\e^{\iu\delta_0}\Big(1+\frac{\epm \omega_0}{\vgx}\chi_t\Big)\e^{\iu k_x x} \um.
\label{eq:xip}
\ee

\subsection{The $x\to-\infty$ asymptotics}

For $x\to-\infty$ we can substitute the Green's function by the corresponding asymptotic
Green's function, given by equation (\ref{eq:gfan}), and we can neglect the contribution to the
integral in $dx^\prime$ of the region in which $|x^\prime|$ is of the order of, or larger than, $|x|$, since
 $V$ is a localized operator.
Using the asymptotic form of $\phi_{k_x}(x)$ and $G_{\mathrm{as}}^+$ given by equation (\ref{eq:gfan}),
we get for $x\to-\infty$
\be
\xi_{k_x}^+(x) \sim -\iu\e^{-\iu\delta_0}\e^{\iu k_x x} \um + \frac{\iu\epm\omega_0}{\vgx}\e^{\iu\delta_0}\e^{-\iu k_x x}\Pm
\mathcal{R} \um,
\ee
where the $2\times 2$ matrix $\mathcal{R}$ depends only on $k_x$ and $k_z$ and is given by
\be
\mathcal{R} = \int_{-\infty}^\infty dx^\prime \phi_{k_x}(x^\prime)
\int_{-\infty}^\infty dx^{\prime\prime}V(x^\prime,x^{\prime\prime}) \phi_{k_x}(x^{\prime\prime}).
\ee
Taking into account the form of $V(x^\prime,x^{\prime\prime})$, the matrix elements $r_{ij}$ of $\mathcal{R}$ are
given by
\be
r_{ij} = \int dx^\prime \int dx^{\prime\prime} \phi_{k_x}(x^\prime) f_{ij}(x^\prime,x^{\prime\prime},k_z)\phi_{k_x}(x^{\prime\prime}),
\label{eq:rij}
\ee
with the functions $f_{ij}$ defined by equations (\ref{eq:t11})-(\ref{eq:t22}).

Since $\Pm$ projects onto the subspace spanned by $\um$, we have
$\Pm\mathcal{R}\um=\chi_r\um$, where $\chi_r$ is a complex number that can be computed
in terms of the $r_{ij}$:
\be
\chi_r = \frac{1}{2}\big(r_{11}+r_{22}\big) + \frac{\iu}{2}\big(\sqrt{\omega_1/\omega_2}r_{12}
-\sqrt{\omega_2/\omega_1}r_{21}\big).
\ee
Hence we have that for $x\to-\infty$
\be
\xi_{k_x}^+(x) \sim -\iu\e^{-\iu\delta_0}\e^{\iu k_x x} \um + \frac{\iu\epm\omega_0}{\vgx}\chi_r\e^{\iu\delta_0}\e^{-\iu k_x x} \um,
\label{eq:xim}
\ee

\subsection{The scattering parameters}

Inspecting equations (\ref{eq:xim}) and (\ref{eq:xip}) we see that by multiplying $\xi_{k_x}^+$ by
$\iu\e^{\iu\delta_0}$ we get the asymptotic behavior
\be
\xi_{k_x}^+(x) \sim \e^{\iu k_x x} + R\,\e^{-\iu k_xx}, \quad \xi_{k_x}^+(x) \sim T\,\e^{\iu k_x x},
\ee
for $x\to -\infty$ and $x\to+\infty$, respectively, where
\be
R = -\frac{\epm\omega_0}{\vgx}\e^{\iu 2\delta_0}\chi_r, \quad T = \e^{\iu 2\delta_0}\Big(1-\frac{\epm\omega_0}{\vgx}\chi_t\Big),
\ee
are the reflection and transmission amplitudes, respectively.
Thus, in the Born approximation the reflection coefficient is
\be
|R|=\epm \frac{\omega_0}{\vgx}|\chi_r|,
\ee
while the transmission
coefficient is, to this order of approximation, $|T|=1$, since $\chi_t$ is purely imaginary.
The reflected and transmitted waves pick up phases,
$\varphi_r$ and $\varphi_t$, respectively, with respect to the incident wave, which are given by
\be
\varphi_r = 2\delta_0 + \delta\varphi_r, \quad \varphi_t = 2\delta_0 + \epm \delta\varphi_t,
\label{eq:dphi}
\ee
where
\be
\delta\varphi_r = \pi + \atan\Big(
\frac{\sqrt{\omega_1/\omega_2}r_{12}-\sqrt{\omega_2/\omega_1}r_{21}}{r_{11}+r_{22}}\Big), 
\ee
\be
\delta\varphi_t = 
\frac{\omega_0}{2\vgx}\big(\sqrt{\omega_1/\omega_2}\,t_{12}-\sqrt{\omega_2/\omega_1}\,t_{21}\big).
\ee
The dependence of the phases on the wave vector originates a shift of the center of the scattered wave
packets, with respect to the center of the incident wave packet, given by
\be
\delta x_l = -\frac{\partial\varphi_l}{\partial k_x}, \quad
\delta z_l = -\frac{\partial\varphi_l}{\partial k_z},
\ee
where the subscript $l$ stands either for $r$ (reflected) or for $t$ (transmitted). These relations are
obtained from a stationary phase analysis, and
imply that the scattered waves propagate along lines shifted laterally with respect to the
prediction of the geometrical optics limit by an amount given by
\bea
&&\delta s_r =
\sin\alpha\frac{\partial\varphi_r}{\partial k_x} + \cos\alpha\frac{\partial\varphi_r}{\partial k_z},
\\[2pt]
&&\delta s_t =
\sin\alpha\frac{\partial\varphi_t}{\partial k_x} - \cos\alpha\frac{\partial\varphi_t}{\partial k_z},
\eea
where $\alpha$ is the incidence angle: $\alpha=\atan(v_z/v_x)$,
with $v_z=\partial\omega/\partial k_z$.

The reflection coefficient vanishes at normal incidence ($k_z=0$), as can be seen by a careful analysis
of the integrals (\ref{eq:rij}), and thus in this case all the energy carried by the
spin wave is transmitted. The contribution of the dipolar interaction to the transmitted amplitude 
also vanishes at normal incidence, as can be seen from equations (\ref{eq:t12poly}) and (\ref{eq:t21poly}).
It is however interesting that even in this case the transmitted wave is shifted
laterally from the incidence direction, since $\partial t_{21}/\partial k_z$ does not vanish at $k_z=0$.
To order $\epsilon$, the amount of the lateral shift at normal incidence is given by
\be
\begin{split}
\delta s_t = \frac{\epm}{4k_x}\frac{q_0^4}{k_x^2+q_0^2}\left[\frac{2}{(\iu k_x-4q_0)^2} + 
  \frac{2}{(\iu k_x-2q_0)^2} \right. \\  \left.
  - \frac{1}{k_x^2} - \frac{1}{2q_0^2}\psi^\prime\Big(\frac{\iu k_x-4q_0}{2q_0}\Big)
      + c.c.\right].
\end{split}
\ee
Clearly, by symmetry the shift at normal incidence should vanish in the case of a 360$\dg$ wall, and indeed
it has been shown that it does vanish in the case of the chiral soliton of monoaxial
helimagnets \cite{Laliena2021}. But symmetry is absent in the case of the 180$\dg$ domain wall considered
here, since the spin wave propagates between two domains with opposite magnetizations. Hence, as we have seen,
the lateral shift does not vanish at normal incidence. It has the direction of the
magnetization of the domain to which the wave is transmitted (in our coordinate system, the $-\vz$ direction).

\begin{figure}[t!]
\centering
\includegraphics[width=0.49\linewidth,angle=0]{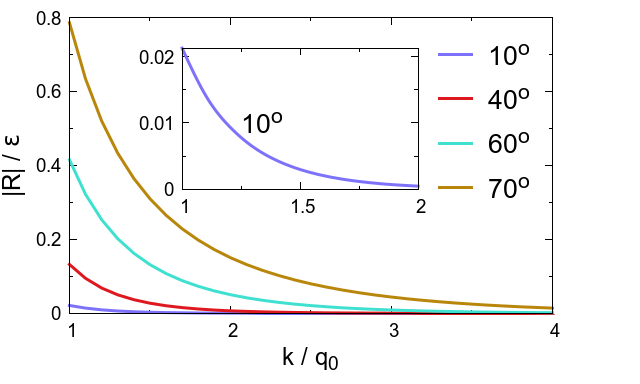}
\includegraphics[width=0.49\linewidth,angle=0]{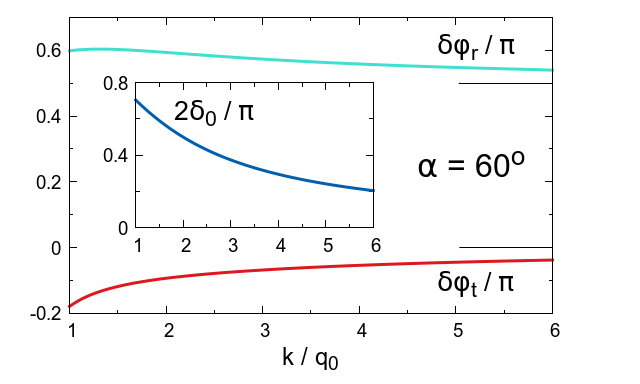}
\caption{Reflection coefficient (left) for the incidence angles indicated in the legend and phases of the scattered
  waves (right) for incidence angle $\alpha=60\dg$. 
\label{fig:Rp}}
\end{figure}

\subsection{Some results}

Let us discuss some results obtained by numerical evaluation of the integrals (\ref{eq:rij}),
and of the right-hand-side of equations (\ref{eq:t12poly}) and (\ref{eq:t21poly}) given in appendix C.

Let $k_x=k\cos\alpha$ and $k_z=k\sin\alpha$,  where $k$ is the modulus of the wave vector and $\alpha$ the incidence
angle. Figure~\ref{fig:Rp} (left) displays the reflection coefficient $|R|/\epm$ as a function of $k$ for
several values of the incidence angle. Actually, we plot the limit $\epm\to0$ of $|R|/\epm$, since we consider
$\epm$ small. As discussed in the previous section, the reflection coeficient vanishes at normal incidence
($\alpha=0$). We see from Figure~\ref{fig:Rp} (left) that $|R|/\epm$ is very small for $\alpha=10\dg$.

Figure \ref{fig:Rp} (right) displays the phases of the scattered waves induced by the dipolar interaction,
$\delta\varphi_r$ and $\delta\varphi_t$, for incidence angle $\alpha=60\dg$.
The inset shows the phase shift in absence of dipolar interaction, $2\delta_0$.
Again, we keep only the first order in $\epm$ and consequently we set $\epm=0$ in $\delta\varphi_r$ and
$\delta\varphi_t$ [see the definitions (\ref{eq:dphi})].
We see that $\varphi_r\to\pi/2$ as $k\to\infty$ (since $\delta_0\to0$), as it happens to be usually for reflected
waves. Analogously, we see that $\varphi_t\to 0$ as $k\to\infty$, as it is expected for a transmitted wave.

\begin{figure}[t!]
\centering
\includegraphics[width=0.49\linewidth,angle=0]{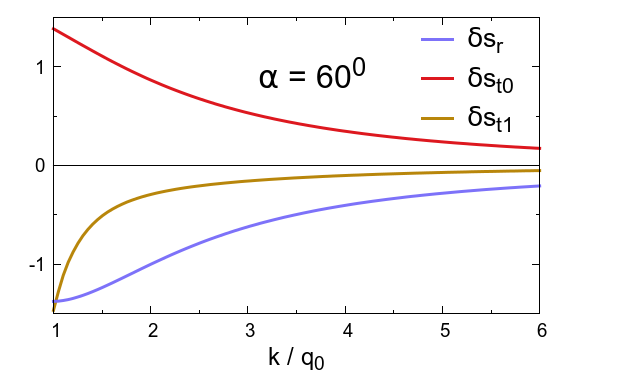}
\includegraphics[width=0.49\linewidth,angle=0]{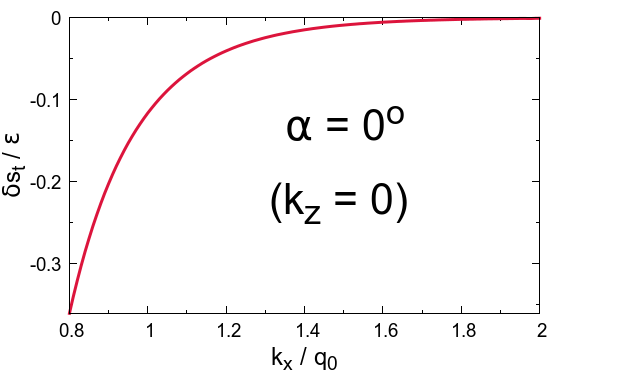}
\caption{Lateral shift of the scattered waves at incidence angle $\alpha=60\dg$ (left) and at normal incidence,
  $\alpha=0\dg$ (right), in units of the domain wall width, $1/q_0$. 
\label{fig:shift}}
\end{figure}

The left panel of Figure \ref{fig:shift} shows the lateral shift of the scattered waves, in units of the domain wall
width, $1/q_0$, as a function of the wave number $k$ for $\alpha=60\dg$. To avoid choosing a particular value of $\epm$,
we use equation (\ref{eq:dphi}) to split the shift of the transmitted wave as
$\delta s_t = \delta s_{t0} + \epm\delta s_{t1}$, where $\delta s_{t0}$ comes from the $2\delta_0$ contribution to
$\varphi_t$ and $\delta s_{t1}$ comes from the $\delta\varphi_t$ term of $\varphi_t$.
The figure displays $\delta s_{t0}$ and $\delta s_{t1}$, with $\epm=0$ in this last quantity.
The two terms have opposite signs and therefore tend to cancel, but the degree of cancellation depends on $\epm$.
It is seen that the shifts are of the order of the wavelength for $k$ of the
order of $q_0$ (i.e., for wavelengths of the order of the domain wall width), and vanish as $k\to\infty$, as expected.
If the reflection coefficient is small enough, and this depends on the actual value of $\epm$, the shift of the
transmitted waves can be enhanced by making the spin wave propagate through an array of well separated domain walls,
since the shift is clearly additive.

The lateral shift for normal incidence, in units of $1/q_0$, is shown as a function of the wave number, $k_x$,
in Figure \ref{fig:shift} (right). Again, to avoid choosing a value for $\epm$ we actually plot the limit $\epm\to0$ of
$\delta s_t/\epm$. We see that it is negative, what means that the transmitted wave at normal incidence is shifted
laterally towards the direction of the magnetization of the domain into which the spin wave is transmitted (the $-\vz$
direction in our coordinate system).
The shift decreases with the wave number, and it is a fraction of the wavelength. Its actual size is proportional to
$\epm$. Given that the reflection coefficient vanish at normal incidence, the lateral shift of the transmitted wave
might be greatly enhanced by using an array of well separated domain walls. The existence of this shift may be an
interesting tool to control and manipulate the spin waves.


\section{Conclusions}
\label{sec:conc}

If the dipolar interaction is neglected, or if it is approximated by a local effective anisotropy field,
the theoretical computations show that a Bloch domain wall of an anisotropic ferromagnet is transparent to
spin waves \cite{Braun1994,Winter1961,Thiele1973}.
However, we have shown in this paper that if the dipolar interaction is taken into account properly
the spin waves are actually reflected by a Bloch domain wall.
The scattering parameters have been obtained perturbatively,
using the distorted wave Born approximation. The application of this perturbative tecnique is not straightforward,
due to the non localized character of the dipolar interaction. It is necessary to split the dipolar contribution
to the spin wave operator into two terms: an operator that can be absorbed into the term treated exactly and an
operator which is localized and can be treated perturbatively in the first (distorted wave) Born approximation.

The scattering parameters can be computed within this distorted wave Born approximation. It turns out that the
reflection coefficient vanishes \textit{only} for normal incidence.
The phase shifts are different for the transmitted and reflected waves, due to the fact that the wall separates
two domains with opposite magnetization, and therefore the mirror symmetry about the wall plane is broken.
The phase shifts depend not only on the wave vector component
perpendicular to the wall plane but also on the component parallel to the wall plane.
The dependence of the phase shifts on the wave vector induce a lateral shift of the reflected and transmitted
waves.
It is worthwhile to stress that the lateral shift of the transmitted wave does not vanish
at normal incidence, due to the lack of symmetry caused by the reversal of the magnetization between the two
domains separated by the wall. In this case the shift has the direction of the magnetization
of the domain to which the spin wave is transmitted. Since the reflection coefficient vanish at normal incidence,
this shift can be greatly enhanced by forcing the spin wave to go through an array of well separated domain walls. 
These properties of the scattering by a domain wall may be very useful to control the spin waves.


\vspace{0.5truecm}

\noindent
\textsc{Acknowledgements} \\

Grant Number PGC2018099024B100 funded by MCIN/AEI/10.13039/501100011033 supported this work.
Grants OTR02223 from CSIC/MICIN and DGA/M4 from Diputaci\'on General de Arag\'on (Spain) are also acknowledged.


\appendix

\section{}

In this appendix we giev some details on the derivation of the $\tilde{D}_{\alpha\beta}$ which gives the
contribution of the dipolar interaction to the spin wave operator. It is studied in section \ref{sec:df}.
Equation (\ref{eq:df}) shows that the dipolar field $\delta\vh_d$ created by the perturbation $\delta\vn$
has the form of a convolution between the Coulomb potential $\vrv/r^3$ and
\be
\nabla\vdot\delta\vn = \partial_x\xi_1 + \cos\theta(x)\partial_y\xi_2 - \sin\theta(x)\partial_z\xi_2.
\ee
Therefore, in terms of the Fourier transform of $\xi_\alpha$ with respect to $y$ and $z$, given by (\ref{eq:xif}),
the dipolar field at point $\vrv=x\vx+\vrvt$ has the form
\be
\delta\vh_d\!=\!-\!\int\!\frac{d^2\kt}{(2\pi)^2}\e^{\iu\vkt\vdot\vrvt}\!\!\int dx^\prime \boldsymbol{C}(x-x^\prime,\vkt)
M(x^\prime,\vkt),
\label{eq:hdf1}
\ee
where
\be
M(x^\prime,\vkt)=\partial_{x^\prime}\tilde{\xi}_1(x^\prime,\vkt)
+\iu\kt F(x^\prime,\vkt) \tilde{\xi}_2(x^\prime,\vkt),
\ee
with $F(x,\vkt)$ given by equation (\ref{eq:F}), and
\be
\boldsymbol{C}(x,\vkt) =
\frac{1}{4\pi}\int d^2\rt \, \e^{-\iu\vkt\vdot\vrvt} \frac{x\vx + \vrvt}{\big(x^2+\rt^2\big)^{3/2}}.
\ee
The above expression can be written as
\be
\boldsymbol{C}(x,\vkt) = xA(x,\vkt)\vx + \iu\nabla_{\kt}A(x,\vkt), \label{eq:C}
\ee
where
\be
A(x,\vkt) =
\frac{1}{4\pi}\int d^2\rt \, \e^{-\iu\vkt\vdot\vrvt} \frac{1}{\big(x^2+\rt^2\big)^{3/2}}.
\ee
For $x\neq 0$ this integral can be readily performed and is equal to
\be
A(x,\vkt) =
\frac{1}{2}\frac{\e^{-\kt |x|}}{|x|}.
\ee
From this and (\ref{eq:C}) we obtain
\be
\boldsymbol{C}(x,\vkt) = \frac{1}{2}\sigma(x)\e^{-\kt |x|}\vx - \iu \frac{1}{2}\e^{-\kt |x|}\frac{\vkt}{\kt}.
\ee
Inserting the above expression into (\ref{eq:hdf1}) and integrating by parts the term that involves
$\partial_{x^\prime}\tilde{\xi}_1$, taking into account that the boundary term vanishes since $\tilde{\xi}_1$
vanishes for large $|x|$, and that $d\sigma(x)/dx=2\delta(x)$, we get
\be
\delta\vh_d = -\int \frac{d^2\kt}{(2\pi)^2} \,\e^{\iu\vkt\vdot\vrvt}\int dx^\prime
\boldsymbol{\Upsilon}(x,x^\prime,\vkt),
\label{eq:hdf2}
\ee
where
\be
\boldsymbol{\Upsilon}(x,x^\prime,\vkt) = \vx \, \Theta(x,x^\prime,\vkt)
+ \frac{\vkt}{\kt} \,\Phi(x,x^\prime,\vkt), \label{eq:upsilon}
\ee
width
\be
\begin{split}
\Theta(x,x^\prime,\vkt) = \Big(\delta(x-x^\prime)-\rho(x-x^\prime)\Big)\tilde{\xi}_1(x^\prime,\vkt) \\
+ \iu\sigma(x-x^\prime)\rho(x-x^\prime)F(x^\prime,\vkt)\tilde{\xi}_2(x^\prime,\vkt),
\hspace*{0.4cm}
\end{split}
\label{eq:Theta2}
\ee
\be
\begin{split}
\Phi(x,x^\prime,\vkt) =
\iu\sigma(x-x^\prime)\rho(x-x^\prime)\tilde{\xi}_1(x^\prime,\vkt) \\
+ \rho(x-x^\prime) F(x^\prime,\vkt)\tilde{\xi}_2(x^\prime,\vkt). \hspace*{0.8cm}
\end{split}
\label{eq:Phi2}
\ee
From equations (\ref{eq:hdf2})-(\ref{eq:Phi2}) it is straightforward to obtain $\tilde{D}_{\alpha\beta}$
using equations (\ref{eq:Ddef1}) and (\ref{eq:Ddef2}).

\section{}

Let us analyze the pole structure in $k_x^\prime$ of the integrand entering the right hand side of equation
(\ref{eq:gfs}). There are two poles coming from $\phi_{k_x^\prime}(x)\phi_{k_x^\prime}^*(x^\prime)$, located on the
imaginary axis, given by $\pm\iu q_0$ (golden points in Figure \ref{fig:poles}). The contribution of these to
poles to the integral gives a function exponentially decreasing with $|x-x^\prime|$ and thus it vanishes
asymptotically. They do not contribute to the asymptotic part of the Green's function.

Let introduce the variable $z=k_x^{\prime\,2}$. The other poles come from the zeros of
\be
f(z) = \omega_1(z)\omega_2(z)+(-\iu\omega+\mu)^2.
\ee
Let us consider first the case $k_z\neq 0$. Since $z=-k_z^2$ is a pole of $f(z)$, it
is clear that
$p(z)=(z+k_z^2)f(z)$ has the same zeros as
$f(z)$. But $p(z)$ is a polynomial of third degree and therefore it has three roots.
Hence, $f(z)$ has exactly three zeros, which are the solutions of
\be
g(z) = \omega^2 + 2\omega\mu\iu - \mu^2, \label{eq:gzeros}
\ee
where, for convenience, we define $g(z) = \omega_1(z)\omega_2(z)$.

Only the poles which attain the real axis as $\mu\to0^+$ do contribute to the asymptotic behavior of the
Green's function (see section \ref{sec:gf}). This means that we only need the zeros of $f(z)$
which attain the positive real axis as $\mu\to0^+$. Let us set $\mu=0$ in equation (\ref{eq:gzeros}).
We notice two facts: i) $g(0)=\omega_G^2$; and ii) it is straightforward to see that
$g^\prime(z)>0$ for $z\geq 0$, where the prime stands for the derivative.
Therefore the equation $g(z)=\omega^2$ has one and only one solution on the positive real axis
if $\omega\geq\omega_G$, and it has no real positive solution if $\omega<\omega_G$. The other two zeros of
$f(z)$ are either non real or negative in the limit $\mu\to 0^+$.

Let us consider a frequency $\omega\geq\omega_G$
and let us denote by $z=k_x^2$ the unique positive
solution of $g(z)=\omega^2$. For $\mu>0$ and small we can obtain the solution of equation
(\ref{eq:gzeros}) as a power series of $\mu$. To leading order we get
\be
z = k_x^2 + \iu\frac{2\omega\mu}{g^\prime(k_x^2)} + O(\mu^2).
\ee
For $\mu\to 0^+$ the imaginary part of the above expression is positive.
This zero of $f(z)$ gives rise to the two poles that contribute to the asymptotic part of the
Green's function:
\be
k_x^\prime = \pm \left(k_x + \iu\frac{\omega\mu}{k_xg^\prime(k_x^2)}\right).
\ee
One of the poles is located on the upper right quadrant of the complex plane and another one on the
lower left quadrant of the complex plane.

The case $k_z=0$ is simpler, since then $g(z)$ is a polynomial of second degree and its zeros have a
relatively simple explicit expression. For $\mu>0$ and small we obtain the two poles 
\be
k_x^\prime = \pm \left( k_x + \iu\frac{\omega\mu}{k_x[\omega^2-\omega_G^2+\omega_0^2(1+\epm/2)^2]}\right).
\vspace{2pt}
\ee
Again one of the poles is located on the upper right quadrant of the complex plane and another one on the
lower left quadrant of the complex plane. Both attain the real axis as $\mu\to 0^+$.


\section{}

The coefficients $t_{12}$ and $t_{21}$ defined by equations (\ref{eq:tij}), (\ref{eq:t12}), and (\ref{eq:t21})
can be evaluated in terms of the derivative of the digamma function, $\psi^\prime(z)$. Let us remember that the
digamma function, $\psi(z)$, is the derivative of the logarithm of the Gamma function.
Defining the complex variable $\lambda=(|k_z|+\iu k_x)/q_0$, the explicit expressions are
\begin{widetext}
\bea
t_{12} &=& -\frac{2k_z^2}{q_0\left(q_0^2+k_x^2\right)}
\left[
1+\frac{k_z}{q_0}\left(
\frac{1}{( \lambda-3)^2} +\frac{1}{( \lambda-1)^2}
-\frac{1}{4}\psi^\prime\Big(\frac{\lambda-3}{2}\Big)
+c.c.
\right)
\right], \label{eq:t12poly} \\
t_{21} &=& -\frac{k_z}{q_0^2+k_x^2}
\left[
\frac{2}{(\lambda-4)^2}
+\frac{2}{(\lambda-2)^2}
+\frac{q_0+k_z}{q_0\lambda^2}
-\frac{1}{2}\psi^\prime\Big(
\frac{\lambda-4}{2}
\Big)+c.c.
\right]. \label{eq:t21poly}
\eea
\end{widetext}

\bibliographystyle{unsrt}
\bibliography{references_dw}

\end{document}